# Social Data Mining through Distributed Mobile Sensing: A Position Paper

John Gekas, Eurobank Research, Athens, GR

**Abstract.** In this article, we present a distributed framework for collecting and analyzing environmental and location data recorded by human users (carriers) with the use of portable sensors. We demonstrate the data mining analysis potential among the recorded environmental and location variables, as well as the potential for classification analysis of human activities. We recognize that the success of such an experimental framework relies on the adoption rate by its candidate user network; thus, we have built our experimental prototype on top of hardware equipment already embedded within the potential users' everyday routine - i.e. hardware sensors installed on modern mobile phones. Finally, we present preliminary analysis results on our collected data sample, as well as potential further work directions and proposed use case scenarios.

## 1      Introduction

The collection, classification and exploratory analysis of social data has been a prolific area of Computer Science research for a number of years. Conceptually, this research area refers to the collection of large-scale user-originated data from a number of social network sources (e.g. Facebook, Twitter, Amazon recommendations and so on), as well as its algorithmic analysis with the purpose of pattern recognition and prediction analysis [OREILLY 2011], [PRITAL *et al,* 2012]. To this end, researchers and analysts attempt to explore patterns and associations in user-based collective behavior, for both business (for example, predictive marketing [SURMA and FURMANEK, 2011], [SURMA and FURMANEK, 2010]) and research purposes alike.

In this paper, we describe a de-centralized framework of *mobile sensor clients* with the ability to collect, analyze and classify user-based data recorded by mobile phone sensors. Mobile phones, when operated as sensor clients, have access to an extended range of user-specific data, including location, mobility, environmental data and so on. In this sense, they have the ability to provide a whole new extended range of social-based information on behalf of their human carrier, when compared to their digital counterparts as means of social data sources (i.e. social network websites).

Our work also relates to the research areas of *wearable mobile sensors* and their various application areas, such as motion analysis [LORINCZ *et al,* 2009], remote healthcare monitoring [KUMAR *et al,* 2012] and activity recognition [UGULINO *et al,* 2012]. However, most experimental solutions are based on custom-built sensor hardware, making the necessary equipment inaccessible to the everyday user, as well as intrusive to use. Mobile phones on the other hand, are already incorporated into most people's everyday lives; as a result, the required hardware is easily accessible to everyday users, and sensor logging operations are less intrusive.

Recent advances in mobile phone technology have turned modern mobile phones (smartphones) into powerful, mobile computers with the ability to collect data regarding the physical environment they are located in [LANE *et al,* 2010], [KANSAL *et al,* 2007]. Although the exact hardware specifications differ (depending on the manufacturer), most mid-range smartphones come equipped with a variety of hardware sensors, including:
- Luminosity measurement
- Sound levels
- Accelerometer (physical forces applied on the 3-axis plane)
- Location, speed (through GPS)
- Temperature
- Image capturing (through the camera)

Due to their relatively low cost, compact size and weight and built-in compatibility with existing software development platforms (such as the Android OS and JAVA programming language), smartphones have already been used in engineering projects where real-time input data from the surrounding environment is required. Such examples are Phonesats[1] (small-size NASA satellites built on top Goggle Nexus smartphones) and Cellbots[2], an open-source robotics community that uses specially modified smartphones as CPU's for robotics projects.

The rest of this paper is organized as follows: Section 2 describes the overall architecture of the introduced framework, as well as the components it consists of. Section 3 presents some relevant Data Mining use cases and analytics directions, based on the social data collected by the mobile sensor framework. Finally, we conclude this paper by discussing some conclusions and further work directions in Section 4.

## 2    Framework Architecture

The proposed framework distinguishes between data collection and data processing. Sensor logging and data collection takes place client-side, with the use of smartphone devices (*mobile sensor clients*) equipped with the required hardware environmental sensors. Sensor clients themselves are not responsible for the storage, analysis and processing of the data, however. Data processing and analysis take place centrally (server-side), on the *base station*. In this sense, the framework consists of the following main components:

---

[1]    NASA Phonesats - http://www.nasa.gov/home/hqnews/2013/apr/HQ_13-107_Phonesat.html.

[2]    Cellbots: Using Cellphones as Robotic Control Platforms - http://www.cellbots.com/.

(a) Mobile Sensor Clients: a distributed network of mobile phones (smartphones) that log and transmit environmental data through the use of their relevant hardware sensors.

(b) Base Station: Desktop database server accessible from all sensor bots, that retrieves, stores and analyzes the data retrieved by the sensor bots.

The conceptual architecture is shown on Fig.1:

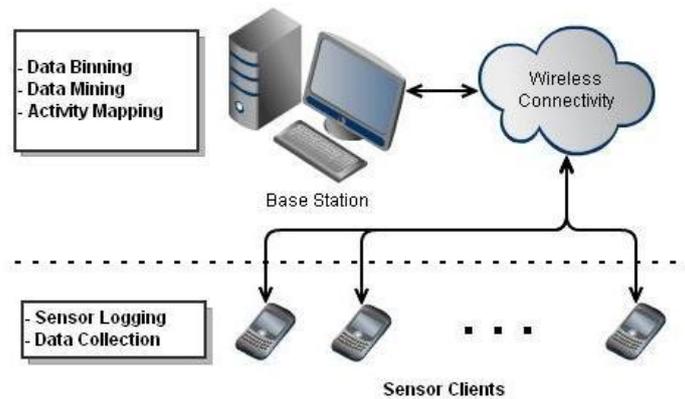

**Fig. 1.** Distributed mobile sensing framework high-level architecture

We will examine the two framework components separately.

### 2.1 Mobile Sensor Clients

*Mobile sensor clients* can be defined as follows: *small, portable computing devices that are attached to their human carriers, and serve the purpose of logging information about their surroundings, as well as transmitting to a central database server for historical storage and further processing.*

The experimental nature of the specific research area dictates that the sensor bots satisfy certain operational requirements: (a) they must have access to various environmental surroundings, (b) mobility, (c) ubiquitous, non-intrusive operation, and (d) connectivity capabilities (for data transmission). In this context, smartphones seem like a perfect candidate for this role. Smartphones, and mobile phones in general already possess significant market share in modern societies, and their role in everyday life is fundamental. As a result, the necessary hardware is already present and distributed among human actors (carriers) and human societies in general.

Regarding the client's technical configuration, we have performed our experiments using a HTC Explorer smartphone, running Android 2.3.1 OS. Furthermore, logging of sensor data and data collection takes place using the AndroSensor 3[rd] party application[3], which is responsible for collecting the input data recorded by the mobile phone's sensors and storing them into locally stored CSV files on the phone's storage memory. Accumulated CSV files are then transferred onto the Base Station wirelessly, once the sensor bot enters the Base Station's vicinity and finds itself within the same Wi-Fi network. The chosen hardware and software configuration is not limiting, however: any mobile phone running a version of the Android OS could be employed for this purpose. Moreover, sensor management and data collection could be performed under an alternative 3$^{rd}$ party application, or a custom-developed one.

---

[3] Androsensor:
https://play.google.com/store/apps/details?id=com.fivasim.androsensor&hl=en

The sensors supported by our hardware configuration are the following:

1. Accelerometer: physical forces measured in m/s$^2$, applied on the mobile phone's 3-axis plane.

2. Luminosity sensor: measures the light intensity of the phone's surroundings.

3. Proximity sensor: examines whether there is another object in close proximity of the mobile phone. It produces a boolean value, indicating whether there is such another object within a 3.5 inch radius.

4. Sound level sensor: measures the sound level (in decibels) of the mobile phone's surroundings.

5. Location sensor: measures the location's longitude, latitude, altitude, as well as the phone's carrier's speed (if any). Measurements are provided through GPS.

The number and variety of the supported sensors depends entirely on the mobile phone's manufacturer, and not on the OS installed on the device or the software application used for sensor management.

### 2.2 Base Station

The *Base Station* is defined as the *central database server, with the purpose of historically storing, processing and analyzing sensor data collected by the mobile sensor clients*.

Technically, it can be perceived as a computer server with wireless connectivity, where accumulated CSV files of recorded sensor data are transferred to by the sensor clients (once they are located within the same wireless network). Transferred input files are then pre-processed and inserted into the database server, where historical data is stored persistently and is available for further processing and analysis. For our experiments, we have used a standard desktop computer running the Windows 7 OS and SQL Server 2005.

The Base Station is responsible for the following operations:
(a) Input of the transferred input files on the database.
(b) Pre-processing and transforming of the input data.
(c) Data binning of continuous sensor variables (for example, sound levels in decibels).
(d) Data mining and analytics operations (such as correlation analysis on separate variables and activity mapping).

The communication flow between the sensor clients and the Base Station is shown on the following flow diagram:

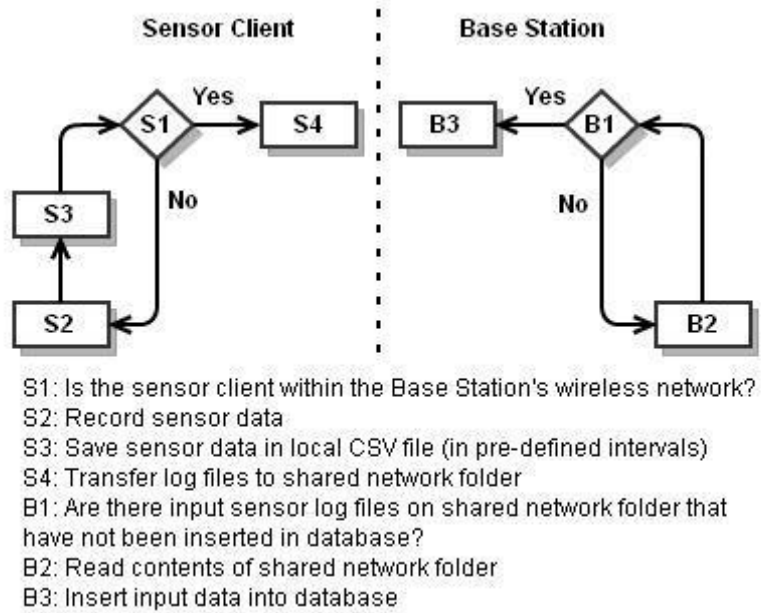

Fig. 2. Communication flow between sensor clients and Base Station

## 3   Social Data Mining - Preliminary Results

The experimental framework described earlier gives us access to environmental (e.g. luminosity, sound levels) and activity-related data (e.g. acceleration forces, speed) coming from a distributed network of human users (device carriers), and thus a wide range of environmental surroundings and activity scenarios. As a result, collected sensor data is highly diverse, both in terms of surroundings and environmental specifications, as well as activities undertaken by different human carriers. In this sense, the Base Station's data repository provides sufficient exploratory ground in order to explore the following research questions:

(a) Is the behaviour of different environmental variables correlated, and to what extent?

(b) Can we infer specific activities undertaken by the users, based on recorded sensor values?

The list of research topics that can be explored is by no means exhaustive; our purpose at this preliminary experimental stage is to present some broad research topics that can be addressed through our experimental framework, rather than a comprehensive list of specific research questions.

The first of the aforementioned research questions can be addressed by applying data analytics techniques such as *statistical correlation* and *association rules learning*, both methodologies popular within the data mining and data analytics communities.

To this end, we have performed preliminary linear correlation analysis among all variables recorded by a single sensor client, based on 3,472 observations collected by the AndroSensor software application. The observations were recorded on 34-second intervals, distributed evenly (following a uniform distribution) between 06:00 in the morning and 00:00 at night. The correlation analysis was performed using the KNIME software tool[4], and its results are displayed on the following Figure, in the form of a correlation matrix:

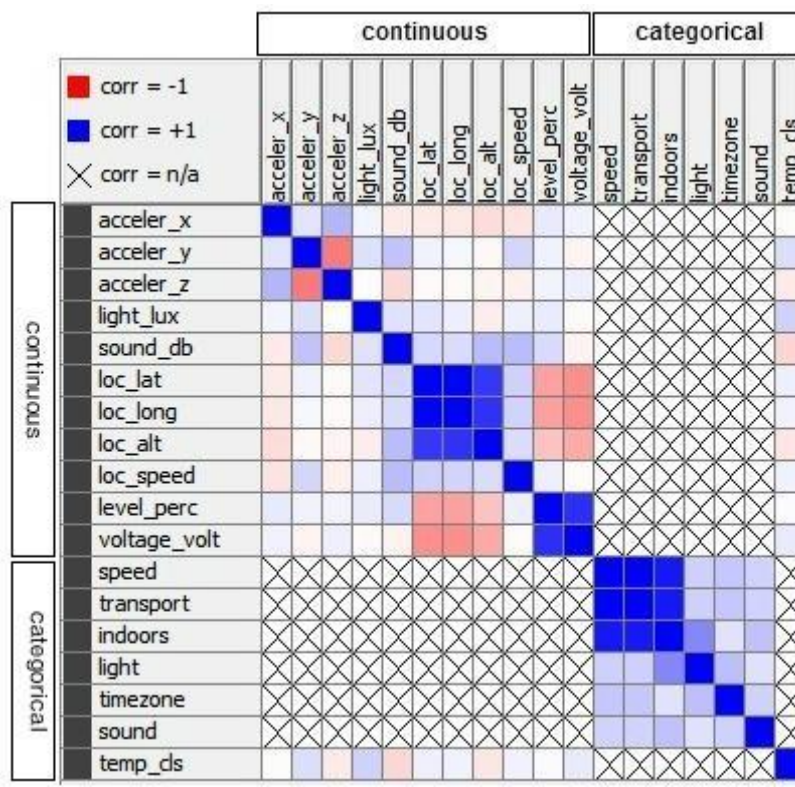

**Fig. 3.** Correlation matrix between recorded sensor variables

Our analysis includes both raw continuous variables, as recorded by the sensor client, and categorical variables produced by data binning on the continuous variables (as marked on the diagram). As we can see on Figure 3, certain variable pairs are positively correlated (at varying correlation degrees), such as speed and sound levels (medium degree), sound levels and luminosity (low degree) and luminosity with indoors/outdoors presence (high degree). On the other hand, there are also variable pairs that are negatively correlated, such as the relationship between the acceleration forces on the Y and Z-axis.

---

[4] KNIME: The Konstanz Information Miner, http://www.knime.org/

## 4 Further Work - Use Case Scenarios

In this paper, we presented the proposed architecture of a distributed mobile sensing framework that provides the ability to collect, store and analyze environmental sensor data recorded collectively by its users. We strongly believe that sensor-recorded data collected within the context of everyday life activities can provide useful insight on analyzing relationships among various environmental variables, as well as recognizing user activities based on repetitive patterns within the recorded parameters. However, data mining analysis on social data requires sufficient data sampling, so that a wide range of user activities and environmental variables are recorded. For this purpose we have employed standard, off-the-shelf mobile phones to operate as our framework's mobile sensing clients: this way, the required equipment is already integrated within most peoples' everyday lives, thus making the process of sensor recording and data collection less intrusive and socially easier to adopt, as well as providing access to a wider range of experimental data.

Our preliminary experimental analysis has examined correlation relationships among the various recorded sensor variables, such as the sensor client's acceleration forces, speed, luminosity and sound level measurements; initial correlation analysis results have been presented in Section 3. In addition, we aim to focus our research towards *human activity recognition*, ie. attempting to infer human activities / actions (eg. *watching television*, *traveling by car or airplane*, *being at a nightclub or the cinema*) through the recorded sensor variables. We believe that the use of classification systems and/or rule-based systems has great potential to this end. Beyond the scientific value of human activity modeling and automatic classification of human actions, we believe the developed technology can also provide added value to social network services. For instance, let us assume that a personal software agent (that operates in collaboration with the mobile sensing framework described in this paper) installed on the user's mobile phone is capable of "knowing" (or inferring) the human carrier's *state*: at home or at work, driving, traveling by train, at a party etc, at real time. This information can then be used for providing extended social network services to the human carrier/user, such as automatic status management on social networking services (eg. Facebook, Twitter), personal recommendation services based on the user's routine and everyday actions, as well as similarity/compatibility analysis among social networks of multiple users, based on real-life activities.